\documentclass[draft,prl,twocolumn,showpacs,groupaddress,preprintnumbers,floatfix]{revtex4-1}

\usepackage{dynlearn}

\newcommand{\qML}{\mbox{quantum-machine}\xspace}
\DeclareMathOperator{\Tr}{Tr}

\begin{document}

\def\ourTitle{%
The Ambiguity of Simplicity
}
%Looking through $\rho$-colored glasses

\def\ourAbstract{%
A system's apparent simplicity depends on whether it is represented
classically or quantally. This is not so surprising, as classical and quantum
physics are descriptive frameworks built on different assumptions that capture,
emphasize, and express different properties and mechanisms. What is surprising
is that, as we demonstrate, simplicity is ambiguous: the \emph{relative}
simplicity between two systems can \emph{change sign} when moving between
classical and quantum descriptions. Thus, notions of absolute physical
simplicity---minimal structure or memory---at best form a partial, not a total,
order. This suggests that appeals to principles of physical simplicity, via
Ockham's Razor or to the ``elegance'' of competing theories, may be
fundamentally subjective, perhaps even beyond the purview of physics itself. It
also raises challenging questions in model selection between classical and
quantum descriptions. Fortunately, experiments are now beginning to probe
measures of simplicity, creating the potential to directly test for ambiguity.
}

\def\ourKeywords{%
  stochastic process, hidden Markov model, \texorpdfstring{\eM}{epsilon-machine}, causal states, mutual information
}

\hypersetup{
  pdfauthor={Cina Aghamohammadi},
  pdftitle={\ourTitle},
  pdfsubject={\ourAbstract},
  pdfkeywords={\ourKeywords},
  pdfproducer={},
  pdfcreator={}
}

\author{Cina Aghamohammadi}
\email{caghamohammadi@ucdavis.edu}
\affiliation{Complexity Sciences Center and Physics Department,
 University of California at Davis, One Shields Avenue, Davis, CA 95616}

\author{John R. Mahoney}
\email{jrmahoney@ucdavis.edu}
\affiliation{Complexity Sciences Center and Physics Department,
 University of California at Davis, One Shields Avenue, Davis, CA 95616}

\author{James P. Crutchfield}
\email{chaos@ucdavis.edu}
\affiliation{Complexity Sciences Center and Physics Department,
University of California at Davis, One Shields Avenue, Davis, CA 95616}

\date{\today}
\bibliographystyle{unsrt}

%%%%%%%%%%%%%%%%%%%%%%%%%%%%%%%%%%%%%%%%%%%%%%%%%%%%%%%%%%%%%%%%%%%%%%%%%%%%%%%
% The paper content

\title{\ourTitle}

\begin{abstract}

\ourAbstract

\vspace{0.1in}

\end{abstract}


\preprint{\sfiwp{16-02-XXX}}
\preprint{\arxiv{1602.XXXXX}}

\title{\ourTitle}
\date{\today}
\maketitle
%\tableofcontents

%\setstretch{1.1}

% \listoffixmes

\epigraph{\emph{We are to admit no more causes of natural things than such as are both true and sufficient to explain their appearances.}}
{Isaac Newton, 1687\\ Philosophi\ae\ Naturalis Principia Mathematica, Book III,
p. 398 \cite{Newt29a}}

\paragraph{Introduction}
Beyond his theory of gravitation, development of the calculus, and pioneering
work in optics, Newton engendered a critical abstract transition that has
resonated down through the centuries, guiding and even accelerating science's
growth: Physics began to perceive the world as one subject to concise
mathematical Laws.  Above, Newton suggests that these Laws are not only a
correct perception (``true and sufficient'') but they are also \emph{simple}
(``admit no more causes''). By his dictates we should abandon the Ptolemaic
epicycle machinery as a description of planetary motion for Newton's more
elegant $F = ma$ and $F_g \propto m_1 m_2/r^2$.

The desire for simplicity in a theory naturally leads us to consider
\emph{simplicity as a means for comparing} alternative theories. Here, we
compare the parsimony of classical physics and quantum mechanics descriptions
of stochastic processes. Classical versus quantum comparisons seem, of late, to
be of much interest both for reasons of principle and of experiment.
\emph{Quantum supremacy} holds that quantum systems behave in ways beyond those
that can be efficiently simulated by classical computers \cite{Pres12a}. In a single cold 2D Fermi gas a spatial transition from core quantum
mechanical states to classical emerges \cite{Boet16a,Fene16a}. And, the
experimental ionization dynamics of highly excited electron states of a single Rydberg atom
are well described by classical chaotic repeller dynamics \cite{Burke11a}. The
impression that one gleans is that it is an interesting time for the
foundations of quantum mechanics. The following adds a new perspective to
these debates on the balance of classical and quantum theory, as concerns the
simplicity of their descriptions.

To start, we consider a Nature full of stationary stochastic processes. A
theory, then, is a mathematical object capable of yielding a process' behaviors
and their probabilities. We can straightforwardly say that one process is more
random than another via comparing their temperatures or their thermodynamic
entropies.  But how to compare them in terms of their structural simplicities?
We make use of a well developed measure of simplicity in stochastic
processes---the statistical complexity \cite{Crut88a}. Measuring a process'
internal memory, it allows for a concrete and interpretable answer to the
question, which process is structurally simpler? Having applied this comparison
to all processes \cite{Feld08a}, we can then lay out the whole space in a neat
array, graded in a linear order from the simplest to the most complicated.

An interesting twist comes about if we add quantum mechanics to our modeling
toolbox. Using descriptions that act on a quantum substrate offers new and
surprising options. For example, it was shown that a quantum mechanical
description can lead to a simpler representation than classical
\cite{Gu12a,Maho15a}. Recently, this quantum advantage was verified
experimentally \cite{Gu16a}. We note in particular that the closed-form methods
introduced in Ref. \cite{Riech15} to measure quantum simplicity obviate many
distracting concerns about generality, approximation, and estimation. This
rigor greatly focuses any ensuing debate on how to measure simplicity.
Leveraging this analysis leads to what is most surprising: what appears to be a
generic quantum simplification is no where near so straightforward. We show that the
\emph{relative simplicity} of classical and quantum descriptions can change.
Specifically, there are stochastic processes, $A$ and $B$, for which the
classical theory says $A$ is simpler than $B$, but quantum mechanics says $B$
is simpler than $A$. What started out as a neat classical array is upended by a
new quantum simplicity order.

To appreciate this, we first discuss in more detail what we mean by
simplicity.  Then, to couch the discussion in terms as physical (and familiar)
as possible, we analyze the one-dimensional Ising spin chain, showing how it
inherently contains such an ambiguity of simplicity. Going further, we
demonstrate that the ambiguity of simplicity is robust: there exist parameter
regions in which the ambiguity is stable against alternative quantum
representations, that arguably would lead to different simplicity metrics. We
show this first for the 2D Ising spin lattice and then establish it generally:
the quantum advantage requires ambiguity. Finally, we draw out potential
impacts for classical-quantum model selection and then propose experimental
tests.

\paragraph{Classical and Quantum Simplicity}
We consider stationary, ergodic processes: each a bi-infinite sequence of random
variables $\MeasSymbol_{-\infty:\infty} = \ldots \MeasSymbol_{-2}
\MeasSymbol_{-1} \MeasSymbol_0 \MeasSymbol_1 \MeasSymbol_2 \ldots$ where each
random variable $\MeasSymbol_t$ (upper case) takes some value $x_t$ (lower
case) in a discrete alphabet set $\MeasAlphabet$ and where all probabilities
$\Pr(\MeasSymbol_{t}, \ldots, \MeasSymbol_{t+L})$ are invariant under time translation.

% We will not explicitly introduce slice notation. Not used enough to
% warrant the space nor the reader confusion.
\renewcommand{\Past}{\MeasSymbol_{-\infty:0}}
\renewcommand{\past}{\meassymbol_{-\infty:0}}
\newcommand{\pastt}{\meassymbol_{-\infty:t}}
\newcommand{\pasttp}{\meassymbol_{-\infty:t^\prime}}
\renewcommand{\Future}{\MeasSymbol_{0:\infty}}
\newcommand{\Futuret}{\MeasSymbol_{t:\infty}}
\newcommand{\Futuretp}{\MeasSymbol_{t^\prime:\infty}}
\renewcommand{\future}{\meassymbol_{0:\infty}}

\emph{How is their degree of randomness quantified?}
Information theory \cite{Cove06a} measures the uncertainty in a single
observation via the \emph{Shannon entropy}: $H[\MeasSymbol_0] = - \sum_{x \in
\MeasAlphabet} \Pr(x) \log_2 \Pr(x)$ and the irreducible uncertainty per
observation via the \emph{entropy rate} \cite{Kolm58}: $\hmu = \lim_{L \to
\infty} H[\MeasSymbol_{0:L}] / L$. If we interpret the left half
$\MeasSymbol_{-\infty:0}$ as the ``past'' and the right half
$\MeasSymbol_{0:\infty}$ as the ``future'', we see that the entropy rate is the
average uncertainty in the next observable given the entire past: $\hmu =
H[\MeasSymbol_0 | \Past]$. Thus, as we take into account the correlations in
the past, the unconditioned single-observation uncertainty $H[\MeasSymbol_0]$
reduces to $\hmu$.

\emph{How reducible is the uncertainty of the future $\MeasSymbol_{0:L}$?}
Naively, this should scale as $L (H[\MeasSymbol_0] - \hmu)$, but due to
correlations within the future, it must be less. The answer comes in the mutual
information between the past and the future, a quantity known as the
\emph{excess entropy} \cite[and references therein]{Crut01a}: $\EE =
\I{\Past : \Future }$. In $\hmu$ and $\EE$, we have measures of
randomness and of how much is predictable in a process, respectively.

Computational mechanics \cite{Crut12a} supplements these with a direct measure
of structure---the amount of process memory. Its main construct, the
\emph{\eM}, is a process's minimal, unifilar predictor. As such, we view a
process' \eM as the ``theory'' of a process: a mechanism that exactly
simulates a process' behaviors.

The \eM consists of \emph{causal states} $\causalstate \in \CausalStateSet$
defined by an equivalence relation $\sim$ that groups histories, say
$\pastt$ and $\pasttp$, that lead to the same future predictions
$\Pr(\Futuret|\cdot)$: $\pastt \sim \pasttp \iff \Prob(\Futuret | \pastt) =
\Prob(\Futuretp | \pasttp)$. In other words, if a simpler set of states is
sufficient, then that set will be the preferred representation. And so, we see
that the \eM is, in a well defined sense, a process' simplest theory.

Translating this notion of simplicity into a measurable quantity, we ask:
\emph{What is the minimum memory necessary to \emph{implement} the maximal
reduction of future uncertainty (by $\EE$ bits)?} The answer is explicit
when phrased in terms of the \eM: the historical information stored in the
present. Quantitatively, this is the Shannon entropy of the causal state
stationary distribution---the \emph{statistical complexity}:\begin{align}\label{CMUEQ}
  \Cmu =  \H\CausalStateSet
  = -\sum \limits_{\causalstate \in \CausalStateSet}
  \pi_\causalstate \log_2 \pi_\causalstate
  ~,
\end{align}
where $\pi_\causalstate$ is the probability of causal state $\causalstate$.

It is well known that the excess entropy is a lower-bound on the information
size of the \eM: $\EE \leq \Cmu$.
In fact, this relation is only rarely an equality \cite{ryan15a}.
So, while $\EE$ quantifies the amount to which a process is subject to
explanation by an \eM theory, this simplest theory is typically larger,
informationally speaking, than the predictability benefit it confers.
That said, the \eM is the best (simplest) theory.
Thus, we use $\Cmu$ to define our notion of classical simplicity.
It provides an interpretable ordering of processes---process $A$ is simpler than process $B$ when $\Cmu^A < \Cmu^B$.

We may also consider the recently proposed \qML representation of processes \cite{Maho15a, Riech15, Gu12a}.
%Specifically, the set of causal states may be re-represented by a set of pure quantum states whose measurement yields the following stochastic symbol sequence.
The \qML consists of a set $\{ \ket{\eta_k(L)} \}$ of pure \emph{signal
states} that are in one-to-one correspondence with the classical causal
states $\causalstate_k \in \CausalStateSet$. Each signal state $\ket{\eta_k(L)}$ 
encodes the set of length-$L$ words that may follow $\causalstate_k$, as
well as each corresponding conditional probability used for prediction from
$\causalstate_k$. Fixing $L$, we construct quantum states of the form:
\begin{align}
\label{eq:qML_def}
\ket{\eta_j(L)} \equiv
  \sum \limits_{w^L \in |\MeasAlphabet|^L}
  \sum \limits_{\causalstate_k \in \CausalStateSet}
  {\sqrt{\Prob(w^L, \causalstate_k | \causalstate_j)}
  ~ \ket{w^L} \ket{\causalstate_k}}
  ~,
\end{align}
where $w^L$ denotes a length-$L$ word and $\Prob(w^L,\causalstate_k |
\causalstate_j) = \Prob(X_{0:L} = w^L,\CausalState_L = \causalstate_k |
\CausalState_0 = \causalstate_j)$. Due to \eM\ unifilarity, a word $w^L$
following a causal state $\causalstate_j$ leads to only one subsequent causal
state. Thus, $\Prob(w^L, \causalstate_k | \causalstate_j) = \Prob(w^L |
\causalstate_j)$.  The resulting Hilbert space is the product $\mathcal{H}_w
\otimes \mathcal{H}_{\sigma}$.  Factor space $\mathcal{H}_{\sigma}$ is of size
$|\CausalStateSet|$, the number of classical causal states, with basis elements
$\ket{\causalstate_k}$. Factor space $\mathcal{H}_{w}$ is of size
$|\mathcal{A}|^L$, the number of length-$L$ words, with basis elements
$\ket{w^L} = \ket{x_0} \cdots \ket{x_{L-1}}$.

The quantum measure of memory analogous to $\Cmu$ is the von Neumann entropy of the stationary state:
\begin{align}
C_q = - \Tr{\rho \log \rho}~,
\end{align}
where $\rho = \sum_i {\pi_i \ket{\eta_i} \bra{\eta_i}}$. This quantum accounting
of memory is generically less than the classical: $C_q \leq \Cmu$. Moreover, as
with classical representations, the excess entropy provides a lower bound: $\EE
\leq C_q$, due to the Holevo bound \cite{Hole73,Gu12a}. In fact, though rare in
the space of processes, the classical and quantum informational sizes are equal
exactly when both models are ``maximally simple'' or ``ideal'', that is, of
size $\EE$ bits: $C_q = \EE$ and $\Cmu = \EE$.

\newcommand{\Bfield}{b}
\newcommand{\IsingA}{\alpha}
\newcommand{\IsingB}{\gamma}
\newcommand{\IsingC}{\delta}

\paragraph{Ising Chain Simplicity}
To ground these ideas, let us consider the Ising spin chain, familiar from
statistical physics \cite{Brush67}, that historically played a critical role
in understanding phase transitions \cite{Rudolf36}, spin glasses \cite{Edwards75},
and lattice gasses \cite{Lee52}. Its impact has reached well beyond physics,
too, to ecology \cite{Noble15a}, financial economics \cite{Sornet14a}, and
neuroscience \cite{Schneid06}.  Here, we first consider the one-dimensional
nearest-neighbor Ising spin chain in the thermodynamic limit. The Hamiltonian
is given by:
\begin{align}\label{Hamiltonian}
H = - \sum_{<i,j>} (J s_i s_j + \Bfield s_i)~,
\end{align}
where $s_i$, the spin at site $i$, takes on values $\big\{ -1, +1 \big\}$, $J$
is the nearest-neighbor spin coupling constant, and $\Bfield$ is the strength of the externally applied magnetic field.

One can measure each spin in the bi-infinite chain from left to right yielding
the random variables $\ldots X_{-1}, X_{0}, X_{1}  \ldots$. In equilibrium this
defines a stationary stochastic process that has been analyzed using
computational mechanics \cite{Feld97b}. Importantly, spins obey a conditional
independence: $\Prob(\Future | \past) = \Prob(\Future | x_0)$. That is, the
``future'' spins (right half) depend not on the entire past (left half) but
only on the most recent spin $x_0$. Therefore, spin configurations resulting
from the Hamiltonian in Eq.~(\ref{Hamiltonian}) can be modeled by a simple
two-state Markov chain consisting of up ($\uparrow$) and down ($\downarrow$)
states with self-transition probabilities \cite{Feld97b}: $p \equiv
\Pr(\uparrow | \uparrow) = N_{+} / D$ and $q \equiv \Pr(\downarrow |
\downarrow) = N_{-} / D$, where $N_{\pm} = \exp{\beta(J \pm \Bfield)}$ and:
\begin{align*}
 D & = \exp{(\beta J)}\cosh{(\beta \Bfield)}\\
   & \quad +
   \sqrt{\exp{(-2\beta J)} + \exp{(2\beta J)} {\sinh(\beta \Bfield)}^2 }
   ~,
\end{align*}
with $\beta = 1/(k_B T)$.

\begin{figure}
  \centering
\includegraphics[width=\linewidth]{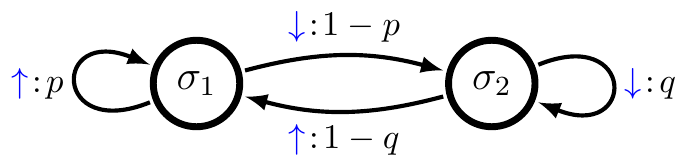}
%  \begin{tikzpicture}[style=vaucanson,
%                      bend angle=15,
%                      scale=1,
%                      every node/.style={transform shape}]
%    \node [state] (A)              {$\causalstate_1$};
%    \node [state] (B) [right of=A] {$\causalstate_2$};
%
%    \path (A) edge [loop left] node {$\Edge{\uparrow}{p}$} (A)
%          (A) edge [bend left] node {$\Edge{\downarrow}{1-p}$} (B)
%          (B) edge [bend left] node {$\Edge{\uparrow}{1-q}$}     (A)
%          (B) edge [loop right] node {$\Edge{\downarrow}{q}$}     (B);
%    % added for symmetry
%    \path (B) edge [loop right, draw=none] node {} (B);
%  \end{tikzpicture}
\caption{The \eM for the nearest-neighbor Ising spin chain has two causal states
  $\causalstate_1$ and $\causalstate_2$. If the last observed spin $x_0$ is up
  ($s_0 = +1$) the current state is $\causalstate_1$ and if it's down ($s_0 =
  -1$) is $\causalstate_2$. If the current state is $\causalstate_1$, with
  probability $p$ the next spin observed is up and, if the current state is
  $\causalstate_2$, with probability $q$ the next spin observed is down.
  }
\label{fig:Ising_eM}
\end{figure}

\begin{figure}
\centering
\includegraphics[width=\linewidth]{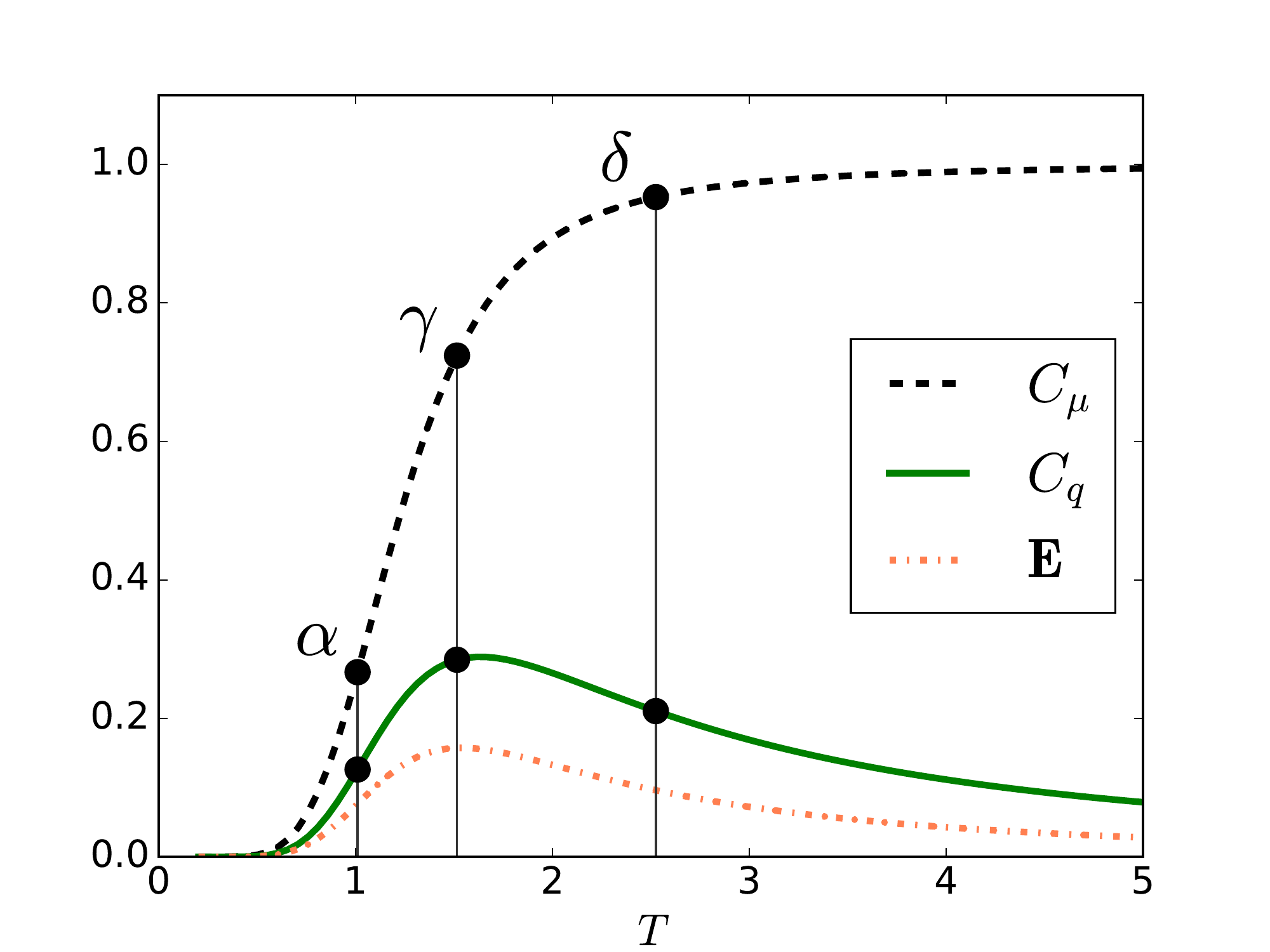}
\caption{Classical and quantum measures of Ising chain simplicity: Statistical
  complexity $\Cmu$, quantum state complexity $C_q$, and excess entropy $\EE$
  versus temperature $T$ in units of ${J}/{k_B}$ at $\Bfield = 0.3$ and $J=1$.
  ($\Cmu(T)$ and $\EE(T)$ after Ref. \cite{Feld98a} and $C_q(T)$ after Ref.
  \cite{Suen15a}.) Three particular spin processes are highlighted $\IsingA$,
  $\IsingB$, and $\IsingC$ at temperatures $T_\IsingA$, $T_\IsingB$, and
  $T_\IsingC$.
  }
\label{fig:Isingfunc}
\end{figure}

Calculating the \eM via the causal-state equivalence relation is
straightforward. There are exactly two causal states; except when $p = 1-q$
where we find only one causal state. The conclusion is that the two-state Markov
chain process is minimally represented by the \eM in Fig.~\ref{fig:Ising_eM}.
Using Eq.~(\ref{CMUEQ}), the statistical complexity is directly calculated as a function of $p$ and $q$.
%\begin{align}
%\Cmu= &- \Big( \frac{1-q}{2-p-q} \Big) \log_2 \Big( \frac{1-q}{2-p-q} \Big) \nonumber \\  
%&- \Big( \frac{1-p}{2-p-q} \Big) \log_2 \Big( \frac{1-p}{2-p-q} \Big)~. 
%\end{align}
Figure~\ref{fig:Isingfunc} shows that $\Cmu$ is a monotonically increasing
function of temperature $T$: $1-\Cmu \propto T^{-2}$ at high $T$.
In particular, for the three processes chosen at temperatures
$T_\IsingA < T_\IsingB < T_\IsingC$: $\Cmu^\IsingA < \Cmu^\IsingB < \Cmu^\IsingC$.

Consider now the quantum representation of the spin configurations. Each causal
state $\causalstate_1$ and $\causalstate_2$ is mapped to a pure quantum state
that resides in a spin one-half space \cite{Suen15a}:
\begin{align}
\nonumber
\ket{\causalstate_1} & = \sqrt{p} \ket{\uparrow} + \sqrt{1-p} \ket{\downarrow}\\
\ket{\causalstate_2} & = \sqrt{1-q} \ket{\uparrow} + \sqrt{q} \ket{\downarrow}~.
\label{Qstates}
\end{align}
Intuitively, the quantum overlap accounts for the fact that the conditional
predictions $\Pr(\Future | \causalstate_1)$ and $\Pr(\Future | \causalstate_2)$
share some subset of future outcomes. The density matrix for the ensemble is
then:
\begin{align}
\rho = \pi_1 \ket{\causalstate_1} \bra{\causalstate_1} + \pi_2 \ket{\causalstate_2} \bra{\causalstate_2}~.
\label{eq:rho_ising}
\end{align}
%where:
%\begin{align*}
%&\rho_{00} = \frac{1-q}{2-p-q} , \quad \rho_{11} = \frac{1-p}{2-p-q}, \\
%&\rho_{01} = \rho_{10} = \frac{(1-q)\sqrt{p(1-p)} + (1-p)\sqrt{q(1-q)} }{2-p-q}~.
%\end{align*}
Computing the quantum analog $C_q = -\Tr{\rho \log \rho}$ as a function of
temperature, Fig.~\ref{fig:Isingfunc} shows that this quantum size is
generically well below the classical size $\Cmu$. Thus, the quantum theory for the
Ising chain is simpler than the classical: $C_q^\IsingA < \Cmu^\IsingA$,
$C_q^\IsingB < \Cmu^\IsingB$, and $C_q^\IsingC < \Cmu^\IsingC$. Given the broad
progress of late in quantum information and computation \cite{Niel10,Pres98a},
it is notable, but perhaps no longer so surprising, that there exists such a
quantum representational advantage.

\paragraph{Ambiguity of Simplicity}
Absolute sizes aside, what can we say about the associated process
\emph{rankings}? How does the notion of ``simpler'' survive the transition from
classical to quantum description?

Observe (Fig.~\ref{fig:Isingfunc}) that, unlike the classical measure $\Cmu$,
the quantum simplicity $C_q$ is not monotonic in the family of processes
reached via increasing temperature: $C_q^\IsingA < C_q^\IsingC < C_q^\IsingB$.
Moreover, the maximum $C_q$ occurs at temperature $T_{C_q} \simeq 1.63$ while
the excess entropy is maximized at temperature $T_{\EE} \simeq 1.53$. Though a
straightforward observation at this point, this basic feature provides the
kernel for drawing out several counterintuitive consequences.

First, what is the consequence of nonmonotonicity itself? Take the processes $\IsingA$
and $\IsingB$ in Fig.~\ref{fig:Isingfunc}. Classically and quantally, $\IsingA$
is simpler than $\IsingB$. In contrast, for processes $\IsingB$ and $\IsingC$
we find that $\IsingB$ is simpler than $\IsingC$ classically, while $\IsingC$
is simpler than $\IsingB$ quantally.

%{\color{red} \jrmnote{Should this be in conclusion? footnote? Now we agree that this should be cut / saved for conclusion but shorter.} 
%One may expect intuitively the quantum and classical models would behave similarly at high temperature, and that the real ``quantum advantage'' would be found at low temperature.
%As we see in Fig.~\ref{fig:Isingfunc} this is certainly not true.
%To understand why, recall that the process is generated by a purely classical system (Ising spin chain).
%The ``quantumness'' of spin systems that we might expect at low temperature is different than the quantum representation of this classical process.
%It would be interesting to investigate the behavior of this classical-versus-quantum analysis of a \emph{quantum} spin chain Hamiltonian (such as the Heisenberg model).}

\begin{figure}
\includegraphics[width=\linewidth]{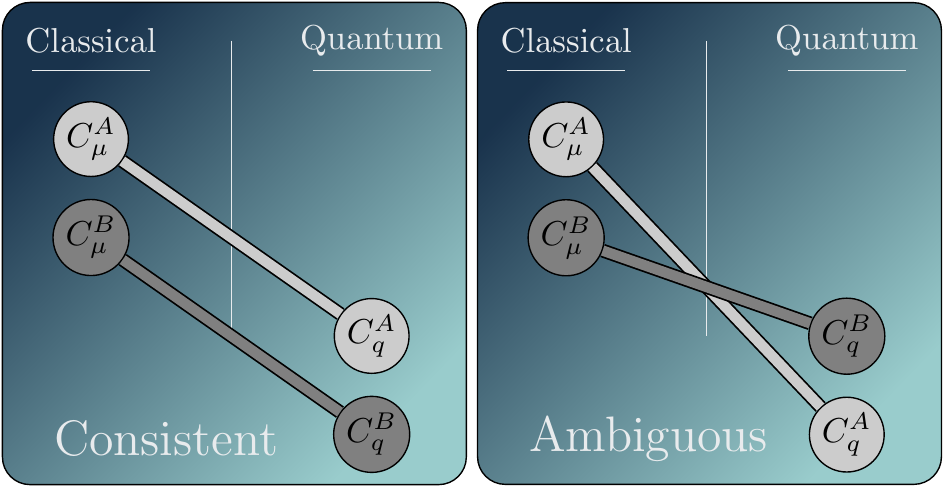}
\caption{(left) Classical and quantum rankings provide a consistent
  interpretation of which process is simpler. (right) Rankings reverse.
  And so, the question of simplicity is ambiguous.
  }
\label{fig:classical_quantum_fancy}
\end{figure}

In this way, even the familiar 1D Ising spin chain illustrates what is a
general phenomenon---the ambiguity of simplicity. How general? Consider two
generic processes $A$ and $B$, for which no change in ranking occurs under the
quantum lens. This indicates a \emph{consistency} between the two
representational viewpoints, at least with respect to processes $A$ and $B$:
$\Cmu^A > \Cmu^B \Leftrightarrow C_q^A > C_q^B$.
Figure~\ref{fig:classical_quantum_fancy}(left) illustrates this circumstance.
Suppose, though, that viewed through our classical lens $B$ appears simpler
than $A$ but, as for the spin chain at high temperature, our quantum lens
reverses the ranking of $A$ and $B$. We refer to this phenomenon as
\emph{ambiguity}. See Fig.~\ref{fig:classical_quantum_fancy}(right). One
concludes that the basic question---``Which process is simpler?''---no longer has a well defined answer.

\begin{figure}
\centering
\includegraphics[width=\linewidth]{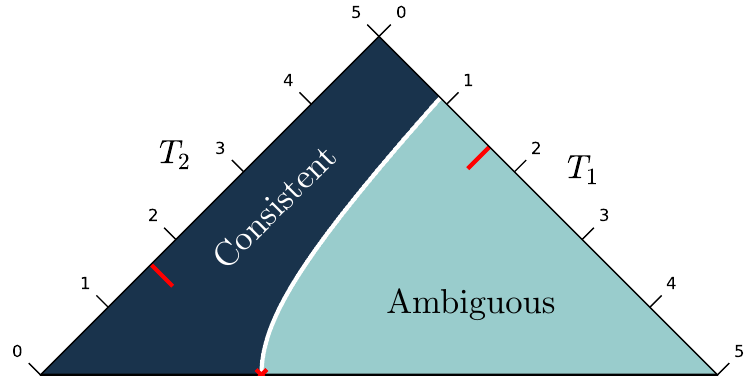}
\caption{Ambiguity diagram for Ising spin chain: Each point corresponds to a
  pair of Ising spin chains at temperatures $T_1$ and $T_2$ with $J=1$ and
  $\Bfield = 0.3$. Consistency is found near the ($T = 0$) axes, while
  ambiguity dominates the remainder of parameter space. Curved boundary between
  these two regions ends at a temperature corresponding to $\max(C_q)$:
  $T_{C_q} \simeq 1.63$ (marked as a red dash).
  }
\label{fig:TvT_triangle}
\end{figure}

How generic are consistency and ambiguity in the Ising spin chain parameter
space? In Fig.~\ref{fig:TvT_triangle} we construct an ambiguity diagram that
compares all pairs of processes at temperatures $T_1$ and $T_2$ in the range
$[0,5]$. There, we fix the magnetic field $\Bfield = 0.3$ and coupling constant
$J=1$. We find that the only consistent pairs are those within a shrinking
envelop around the axes ($T_1 = 0$ and $T_2 = 0$). The bulk of parameter space,
then, contains ambiguously ranked pairs. The singular feature of the diagram is
the leftmost point along the boundary between the two regimes. This occurs at
the temperature $T_{C_q} \simeq 1.63$ where we find the maximum value of
$C_q$.  Monotonicity of $\Cmu$ ensures that a transition from the consistent
region to the ambiguous one is controlled by the reordering of $C_q$ values and
not by $\Cmu$ values.

\paragraph{Robustness of ambiguity}
One can object that this ambiguity is merely an artifact of the particular
quantum model-size measure $C_q$ or of the assumptions in constructing the
quantum states from a process' \eM. This is a valid concern, especially since
minimality of the above \qML representation (or any other quantum
representation, for that matter) has not been established. Critically, as we
now prove, the essence of ambiguity does not depend on this contingency.

Denote by $\widetilde{C_q}$ the memory measure of an optimal quantum model $\widetilde{Q}$ built according to some hypothetical, alternative quantum representational scheme. Since $\widetilde{C_q}$, like $C_q$, is also bounded between $\EE$ and $\Cmu$ (\cite{Hole73, Gu12a}), we can define sufficient criteria for consistency and ambiguity between $\widetilde{C_q}$ and $\Cmu$.
(For the following and without loss of generality, we also assume that the
hypothetical model $\widetilde{Q}$ is at least as efficient as our original
$\qML$: $\widetilde{C_q} \leq C_q$.)

\begin{figure}
\includegraphics[width=\linewidth]{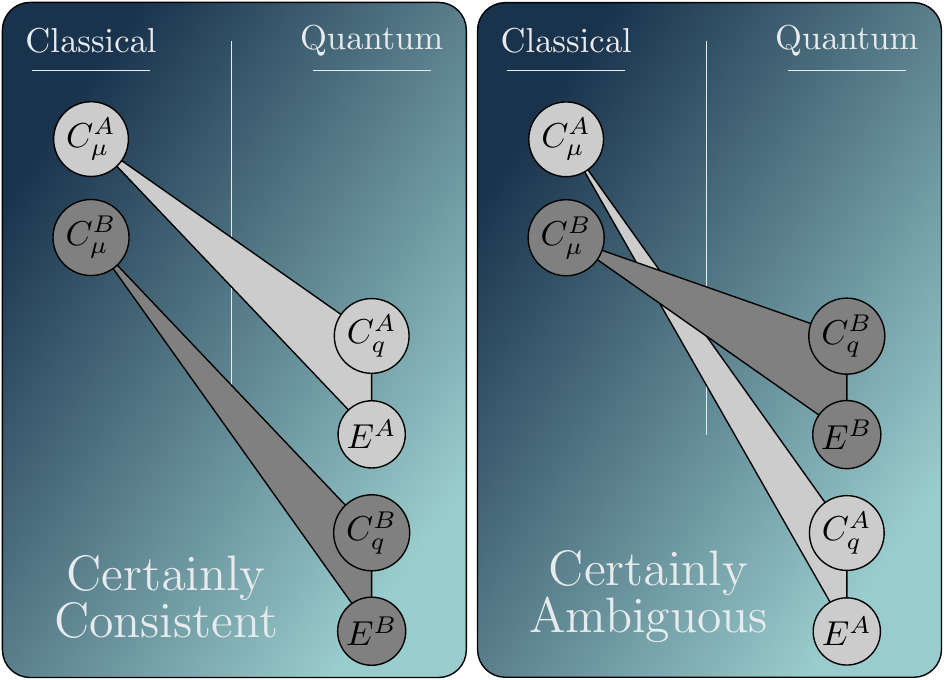}
\caption{Constraining hypothetical, as-yet-unknown frameworks for building   
  quantum models $\widetilde{Q}$: Appealing to size measures $C_q$ and
  $\EE$ and without knowing any further details about $\widetilde{Q}$, we can
  still identify processes for which classical and quantum simplicity orderings
  must \emph{certainly} be consistent or ambiguous. Cases exist that fall into
  neither of these stricter categories.
  }
\label{fig:classical_quantumplus_fancy}
\end{figure}

Assume that for processes $A$ and $B$, $B$ is classically simpler: $\Cmu^B < \Cmu^A$. Then, since $\EE \leq \widetilde{C_q} \leq C_q$, the stronger criterion $\EE^A > C^B_q$ ensures that any $\widetilde{Q}$ must yield consistency in classical and quantum ordering and is therefore, what we call, \emph{certainly consistent}. See Fig.~\ref{fig:classical_quantumplus_fancy}(left). Similarly, if $\EE^B > C^A_q$, we know that any $\widetilde{Q}$ must yield an ambiguous ordering and is \emph{certainly ambiguous}. See Fig.~\ref{fig:classical_quantumplus_fancy}(right).

\begin{figure}
\centering
\includegraphics[width=\linewidth]{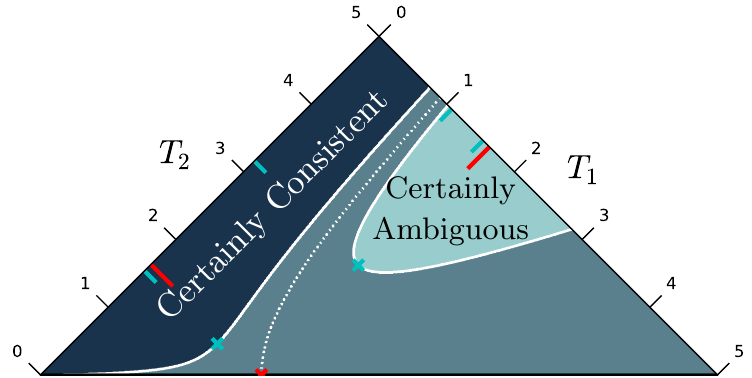}
\caption{Certain ambiguity diagram: Each point corresponds to a pair of Ising
  spin chains at temperatures $T_1$ and $T_2$ with $J=1$ and $\Bfield = 0.3$.
  Dashed line marks previous certainty/ambiguity border of Fig.
  \ref{fig:TvT_triangle}. Certain consistency (ambiguity) is a proper subset of
  consistent (ambiguous). Local extrema of new boundaries at temperatures
  corresponding to $(C_q=\max(\EE), \max(\EE))$ and $(\max(\EE),
  C_q=\max(\EE))$ (marked with short blue lines).
  Long red lines mark the same values as in Fig.~\ref{fig:TvT_triangle}.
  }
\label{fig:TvT_triangle2}
\end{figure}

Figure~\ref{fig:TvT_triangle2} illustrates these stricter relations within the
same Ising parameter region used in Fig.~\ref{fig:TvT_triangle}. The central
region does not satisfy either strict constraint. As expected, the certainly
consistent (ambiguous) area is a proper subset of the consistent (ambiguous)
area.

One concludes that no matter what future improvements may be found in quantum
representations, these ``certain'' subregions are robust and will have known
consistency or ambiguity. This is a strong statement about how one can or
cannot systematically rank the simplicity of systems classically and quantally.
Again, the basic Ising spin chain is sufficiently rich to illustrate these
these new phenomena.

\paragraph{Discussion}
How common is ambiguity? First, what can we say about ambiguity in the
analogous (nearest-neighbor, ferromagnetic) two-dimensional Ising system?
%To answer this question we need to come back and look closely at the important difference between $\Cmu$ and $C_q$.
%While $C_q$ is a smooth function of \eM's transition probabilities, this is not generally true for $\Cmu$.
%Consider Fig.~\ref{fig:Ising_eM} for $p$ and $q$ close to $p=1/2$. 
%In this case, we have a uniform distribution over causal states and consequently $\Cmu \simeq1$.
%For quantum machine, using Eq.~\ref{Qstates}, two states are really close to each other meaning $\braket{\causalstate_1|\causalstate_2}\simeq 1$. The consequence is $C_q \simeq 0$. 
%Now lets look ate the case where $p=q=\half$. 
%The two causal states are not distinguishable and we only have one causal state and as a result $\Cmu=0$. We also have $\ket{\causalstate_1}=\ket{\causalstate_2}$ which leads to $C_q=0$.
%The lesson here is, $C_q$ smoothly tracks how states are distinguishable but $\Cmu$ tracks if that the causal states are completely distinguish or not. 
%Come back to 1D Ising chain, for high temperatures $p\neq q$ but both close to $\half$ and never equal. The consequence is $C_q \simeq 0$ and $\Cmu=1$.
%Now let's look at 2D Ising model. 
At the extreme $T=0$ and for any nonzero value of external field, the ground
state will be in uniform alignment with the field. This means that any random
variable constructed from spin variables must have vanishing entropy. Lacking
a complete computational mechanics of structure in two-dimensional patterns
\footnote{Though see Ref. \cite{Feld98a,Feld02b}.}, it is still clear that any
analog of statistical complexity (and thereby $C_q$) will vanish at $T=0$ for
such uniform configurations.

At very high $T$, though, spins become increasingly uncorrelated and the
probability distribution over configurations approaches uniformity, \emph{but
is not exactly uniform}. That is, for any sufficiently high finite temperature,
the system is has some, perhaps weak, correlation and so is not memoryless.
Causal states in this regime remain probabilistically distinct. So, as with the
1D case, at very high temperature ($T \gg 1$, but $T \neq \infty$) $\Cmu(T)$ is
not zero.

What can we say about $C_q$ in this limit? For high $T \gg 1$ spin randomness
makes the quantum states $\{\ket{\eta}\}$ (Eq.~\ref{eq:qML_def}) more and more
indistinguishable. And so, their increasing overlaps $\braket{\eta_i|\eta_j}
\rightarrow 1$, driving $C_q$ to zero monotonically. The conclusion is that for
the 2D Ising, at $T \approx 0$ and $T \gg 1$,  we have the same qualitative
picture for the simplicity measures as in Fig.~\ref{fig:Isingfunc}. This brief
argument says that ambiguity exists in the 2D Ising spin model as well.

Perhaps the ambiguity of simplicity is special to spin systems. The appendix
shows that it is, in fact, a much more general phenomenon, by introducing a set
of easily satisfied conditions such that two simplicity functions over a set of
structured objects must yield ambiguous ordering. In particular, taking the
space of all \eMs as a set and $\Cmu$ and $\widetilde{C_q}$ as the two measures
, we find that these conditions are satisfied. The general consequence is that either the two
measures selected are trivially equal or ambiguity must exist. In other words, if
the world is not ambiguous, quantum mechanics cannot simplify its explanation.
One concludes that ambiguity is necessary for quantum simplification.

\paragraph{Closing Remarks}
We now see that comparing classical physics and quantum mechanics descriptions
of the world calls into question our basic belief in the simplicity of physical
theories. However, monitoring model simplicity (and therefore model ordering)
is far from being the sole domain of physics. It is key in a variety of
contemporary statistical inference tasks, specifically in model selection
\cite{Burnh03}.

Imagine two competing models $A$ and $B$ of some finite data $\mathcal{D}$.  In
Bayesian inference, one widely employed methodology, choosing one 
over model another requires us to calculate the posterior probabilities that each
generated $\mathcal{D}$. This requires specifying a prior
probability distribution over $A$ and $B$ at the outset \cite{Strel14}. Such priors are
commonly constructed to favor simpler models. Indeed, there is a long history
of methods to avoid overfitting to data that directly incorporate simplicity
measures into model selection, including Akaike's Information Criterion
\cite{Akai74}, Boltzmann Information Criterion \cite{Schw78}, Minimum
Description Length \cite{Riss98a}, and Minimum Message Length \cite{Wall87a}.

Classically, we may find that $A$ is simpler than $B$. This fact then enters
our inference through the model prior, favoring $A$. Given that the two
likelihoods $\Pr(\mathcal{D}|A)$ and $\Pr(\mathcal{D}|B)$ are the same or
similar enough, the inference identifies $A$ as preferred. As we showed, the
tables may turn dramatically when presented with quantum data; we might
find there that $B$ is much simpler. We must then reconcile the fact that had we constructed the
model prior using our quantum lens, $B$ would have yielded as the preferred model.

We introduced the ambiguity of simplicity focusing on classical and quantum
descriptions of classical processes. Quantum supremacy \cite{Pres12a} suggests
we go further to probe how (and if) ambiguity manifests when modeling quantum
processes.  This can be probed in the 1D quantum Heisenberg spin chain
\cite{Bax07}, for example. Measuring each spin within the bi-infinite chain in
the $z$-direction yields a stochastic process---one that can be described
classically or quantally. The Heisenberg spin chain is realized experimentally
in the quasi-1D magnetic order found in antiferromagnetic $KCuF_3$ crystals
\cite{Tenn95a,Tenn95b,Mail07a}. One can then adapt the methods of 1D chaotic
crystallography \cite{Varn14a} to extract the \eM and quantum-machine
descriptions of the quantum crystalline structure from the neutron scattering
measurements. These and perhaps other experiments will provide an entre\'e to
analyzing the ambiguity of simplicity in quantum systems.

\section*{Acknowledgments}
\label{sec:acknowledgments}

We thank A. Aghamohammadi, M. Anvari, R. D'Souza, R. James, F. Pourbabaee, and
D. Varn for useful conversations. JPC thanks the Santa Fe Institute for its
hospitality during visits as an External Faculty member. This material is based
upon work supported by, or in part by, the John Templeton Foundation and the U.
S. Army Research Laboratory and the U. S. Army Research Office under contracts
W911NF-13-1-0390 and W911NF-13-1-0340.

\bibliography{chaos,ref}

%\cleardoublepage

\appendix

\section{Appendix}
\label{sec:appendix}

First, we lay bare the mathematical argument and then we interpret
it in terms of the physical setting of the main text.

Consider a set of objects $S$ and two functions over the set $F_1:S \rightarrow G$ and $F_2:S \rightarrow G$.

If there exists $s_1, s_2 \in S$, such that $F_1(s_1) > F_1(s_2)$ and $F_2(s_1) < F_2(s_2)$, then we say these functions are \emph{ambiguous} over $S$.

We define three conditions for the set and functions.

{\bf Condition A}
The two functions map onto the whole space $G$:
$F_1(S) = G$ and $F_2(S) = G$.

{\bf Condition B}
For all $g \in G$ there exists $x \in S$ such that $F_1(x) = F_2(x) = g$.

{\bf Condition C}
Assume $\preceq$ is a dense, total order on space $G$.

{\The
Given two functions $F_1$ and $F_2$ that map set $S$ to space $G$
and satisfy Conditions A, B, and C: No ambiguity implies that for
all $x \in S$, $F_1(x) = F_2(x)$.
}

{\ProThe
We prove the contrapositive by contradiction.
Assume there exists $x \in S$ such that $F_1(x) \neq F_2(x)$.
Without loss of generality, let $F_1(x) \succeq F_2(x)$.
Since $\preceq$ is a dense total order on $G$, there is $g \in G$
such that $F_1(x) \succeq g \succeq F_2(x)$.
By Condition B, there exists $y \in S$ such that $F_1(y) = F_2(y) = g$.
Trivially then, 
$F_1(x) \succeq F_1(y)$ and $F_2(x) \preceq F_2(y)$.
This demonstrates ambiguity and completes the proof.
}

We can interpret this in the setting of stationary processes with measures
$\Cmu$ and $\widetilde{C_q}$ and discuss the space of all possible quantum
sizes. More specifically, consider the case $F_1 = \Cmu$ and $F_2 =
\widetilde{C_q}$. We know that for any value $y \in \mathbb{R}$, there exists
an \eM\ with $\Cmu = \widetilde{C_q} = \EE = y$. This satisfies the assumption.
Then, our results say that if the world is not ambiguous, the two measures are
equivalent. In other words, the quantum advantage $\widetilde{C_q}$
\emph{requires} ambiguity.

\end{document}